\documentclass[11pt]{article}
\usepackage{acl}
\usepackage{times}
\usepackage{latexsym}
\usepackage{booktabs}
\usepackage{graphicx}
\usepackage{amsmath}
\usepackage{xcolor}
\usepackage{multirow}
\usepackage{url}

\newcommand{\agentsmd}{\texttt{AGENTS.md}}
\newcommand{\none}{\textsc{none}}
\newcommand{\alwayson}{\textsc{always\_on}}
\newcommand{\selective}{\textsc{selective}}

\title{Do Context Files Help Coding Agents?\\A Two-Agent Ablation Study on Real Repositories}

\author{Prakhar Khatri \\ Independent Researcher \\ \texttt{prakharkhatri123@gmail.com}}

\begin{document}
\maketitle

\begin{abstract}
Persistent context files (\agentsmd{}, \texttt{CLAUDE.md}) are standard practice for guiding AI coding agents, yet evidence for their effectiveness is contradictory.
We present a controlled ablation of context-injection strategy across two frontier agents (Claude Code and Codex), 17 real tasks from 3 repositories (15 shared + 2 Codex-only), and 288 evaluated runs with gold-test evaluation.
Context strategy does not measurably move correctness on either agent (bounded to ${\leq}10$--$15$pp via equivalence testing).
A failure-mode triage reveals why: agents fail on \emph{implementation skill}---feature design, pattern selection, exact wiring---not missing repository knowledge that a context file could supply; a manipulation probe confirms the real \agentsmd{} never converts a near-miss to a pass on either agent.
We further show that borderline task difficulty is \emph{agent-specific} (Spearman $\rho{=}0.75$), offering a candidate explanation for prior contradictions: single-agent studies draw tasks from different agents' informative bands.
We release all code, data, and analysis.
\end{abstract}

\section{Introduction}

Repository context files---\agentsmd{}, \texttt{CLAUDE.md}, and similar project-level guides---have become ubiquitous infrastructure for AI coding agents.
Platforms like Claude Code auto-load these files into every session; practitioners invest significant effort authoring coding conventions, architectural constraints, and workflow guidance, expecting agents to produce better code as a result.

Yet empirical evidence is contradictory.
\citet{paper1} report that \agentsmd{} files improve agent \emph{efficiency} (lower wall-clock runtime and fewer output tokens), while \citet{paper2} find no significant effect on task completion when context files are present versus absent.
The two studies differ in agent (Codex-family vs.\ Claude-family), evaluation method, and experimental control---making reconciliation impossible without a study that varies injection strategy under controlled conditions across both agent families.

This paper presents such a study.
We conduct a controlled ablation of context-injection strategy on real merged pull requests, evaluated with hidden gold tests (SWE-bench Tier-C style), across two frontier agents from different providers.
Our three strategies---\none{} (no context), \alwayson{} (full \agentsmd{} in every system prompt), and \selective{} (topic-organized wiki files the agent retrieves on demand)---vary primarily \emph{how} context is delivered (see \S\ref{sec:iv} for a caveat on the \selective{} corpus).

Our contributions are:
\begin{enumerate}
    \item A \textbf{controlled, two-agent ablation} (288 evaluated runs: 15 Claude tasks + 17 Codex tasks, each $\times$ 3 strategies $\times$ 3 repeats) with SWE-bench-style gold-test evaluation on real repositories, yielding a \textbf{bounded null}: the correctness effect is ${\leq}10$pp (Claude) / ${\leq}15$pp (Codex).
    \item A \textbf{failure-mode triage and manipulation probe} demonstrating that real tasks fail on implementation skill, not missing repository knowledge---the mechanism behind the null---and that the real \agentsmd{} cannot convert a near-miss failure to a pass on either agent.
    \item Evidence that \textbf{borderline task difficulty is agent-specific}: the same task may be trivially solvable for one agent yet challenging for another, explaining why single-agent studies reach contradictory conclusions.
\end{enumerate}

\section{Related Work}

\paragraph{Coding agents and benchmarks.}
SWE-bench~\citep{swebench} established the issue-to-PR evaluation paradigm for coding agents, and subsequent work on agent--computer interfaces~\citep{sweagent} shows that harness design (tool ordering, context budgets, retry logic) materially affects performance independent of the underlying model.
Our work builds on this paradigm but studies a different variable: persistent repository context rather than harness mechanics.

\paragraph{Context and memory for agents.}
Persistent memory systems for LLM agents range from operating-system-style memory management~\citep{memgpt} to session-accumulated skill libraries~\citep{voyager}, and agent loops that interleave reasoning, acting, and self-reflection~\citep{react,reflexion}.
A separate line of work shows that \emph{how} long contexts are arranged matters: models use information unevenly across a long prompt, often underusing material placed mid-context~\citep{lostinthemiddle}---directly relevant to whether an always-on \agentsmd{} in the system prompt is actually attended to.
\agentsmd{} files represent the simplest form: a static document loaded at session start.
\citet{paper1} study the impact of these files on Codex-family agents, measuring wall-clock time and token usage; they report efficiency improvements but do not control for injection strategy.
\citet{paper2} run a larger naturalistic study with Claude-family agents and find no significant effect on correctness.
Neither study isolates injection strategy as an independent variable, nor do they test across agent families.

\paragraph{Repository-level context for code generation.}
Retrieval-augmented generation (RAG)~\citep{rag} has been applied to code tasks by retrieving relevant files, functions, or documentation before generation, and repository-level benchmarks and retrieval methods---\textsc{RepoCoder}~\citep{repocoder}, \textsc{CrossCodeEval}~\citep{crosscodeeval}, and \textsc{RepoBench}~\citep{repobench}---specifically target cross-file and repository-scale context.
Recent work on SWE-bench Verified~\citep{swebenchverified} refines the benchmark with human-validated instances, raising the evaluation bar.
However, these efforts focus on \emph{what} context to retrieve from the codebase at query time, whereas \agentsmd{} files represent a \emph{static, author-curated} context layer orthogonal to retrieval---our study isolates whether this layer helps, and how.

\paragraph{Equivalence testing in empirical SE.}
Traditional null-hypothesis testing cannot confirm the absence of an effect~\citep{tost}.
Two one-sided tests (TOST) provide bounded equivalence claims, but remain rare in agent evaluation.
We adopt TOST alongside permutation tests and power analysis to characterize our null, following recommendations from empirical software engineering methodology~\citep{empiricalse}.

\section{Method}

\subsection{Overview}

We measure whether the \emph{strategy} by which repository context is injected into a coding agent affects task correctness or efficiency.
The independent variable has three levels (\none{}, \alwayson{}, \selective{}); the dependent variables are binary correctness (gold tests pass/fail) and continuous efficiency metrics (tool calls, wall-clock time, output tokens, cache usage).
Each task is run under all three strategies with 3 independent repeats, on two agents, yielding a within-task paired design.
In total the design comprises 291 completed agent runs; 3 Claude runs failed to produce a valid gold-test evaluation (crash/timeout) and are excluded from correctness, leaving 288 evaluated cells (efficiency metrics use all completed runs, so their per-cell $n$ varies slightly).
Figure~\ref{fig:strategies} illustrates the three injection strategies.

\begin{figure}[t]
\centering
\small
\setlength{\tabcolsep}{4pt}
\renewcommand{\arraystretch}{1.25}
\begin{tabular}{@{}lccc@{}}
\toprule
& \textbf{File in} & \textbf{In system} & \textbf{Retrievable} \\
\textbf{Strategy} & \textbf{workspace} & \textbf{prompt} & \textbf{wiki files} \\
\midrule
\none      & --- & --- & --- \\
\alwayson  & --- & full \agentsmd{} & --- \\
\selective & --- & retrieval hint & \texttt{wiki/*.md} \\
\bottomrule
\end{tabular}
\caption{The three context-injection strategies. In every condition the workspace \agentsmd{} is removed, so the injection channel is the \emph{only} context the agent receives. \none{} provides no context; \alwayson{} injects the full file each turn; \selective{} places topic-organized wiki files in the workspace that the agent reads on demand, cued by a system-prompt hint. The \selective{} corpus equals the \agentsmd{} for one repository (opshin) but is a broader auto-generated wiki for the others (\S\ref{sec:iv}).}
\label{fig:strategies}
\end{figure}

\subsection{Task Selection and Evaluation}

\paragraph{Repositories.}
We surveyed ${\sim}$40 repositories drawn from three sources: our own manual scan, and the repository sets of \citet{paper1} and \citet{paper2}.
We excluded SWE-bench repositories (none maintain \agentsmd{} files) and filtered candidates on four criteria: (1)~exactly one root \agentsmd{} with no competing instruction stack; (2)~file quality rated Good or Excellent on a structured rubric (coverage of build/test/style conventions, architectural guidance, specificity); (3)~feasible pilot setup (no oversized monorepos); (4)~Python-only, to avoid confounding context effects with build-system or language differences.
Six repositories survived; we narrowed to three for budget feasibility:
\textbf{pdm} (package manager, 477-word context file),
\textbf{firebase-admin-python} (cloud SDK, 1236-word file rated ``Excellent''), and
\textbf{opshin} (domain-specific compiler, 248-word file).
These span distinct domains (tooling, cloud infrastructure, compilers) and \agentsmd{} quality levels (248--1236 words).

\paragraph{Tasks.}
We mine tasks from merged pull requests via the GitHub API: the PR description becomes the agent prompt, the base commit the starting state, and the PR's own test files the gold evaluation.
We keep PRs that (1) add or modify test files (required for gold-test scoring; without tests one falls back to fuzzy diff-overlap, which inflates apparent variance), (2) are non-trivial (multiple files or $>$50 LOC), and (3) need no external credentials or services (for the egress-locked pod).
This yields an initial batch of 12 tasks (4 per repository, spanning medium and complex difficulty).
To avoid a floor/ceiling design---where every task is too easy or too hard for any manipulation to register---we ran a Codex screening sweep over 84 candidate tasks (3 strategies $\times$ 1 repeat) to locate the \emph{borderline} band ($0{<}\text{pass}{<}1$), adding 4 borderline tasks.
Screening also surfaced a portability bug in our effort classifier (\S\ref{sec:classifier}).
The final set is 17 tasks for Codex and 15 for Claude; two high-effort tasks are Codex-only, dropped from the Claude arm under an Anthropic budget cap.
Because the borderline band was calibrated on Codex, the screened tasks need not be borderline for Claude---an instance of the agent-specific difficulty we analyze in \S\ref{sec:borderline}.

\paragraph{Evaluation (Tier-C).}
After the agent completes, we extract test files from the gold PR diff, apply them onto the agent's workspace, and run them against the agent's code.
The task passes if and only if all gold tests pass.
This follows the SWE-bench evaluation protocol: the agent never sees the test files; they serve as an oracle.

\paragraph{Safety and isolation.}
All runs execute on an egress-locked pod with GitHub DNS blackholed.
Git remotes are scrubbed, credentials are stripped, and git push/commit commands are denied via PATH shims.
Future commit history is pruned so the agent cannot read gold solutions from \texttt{git log}.

\subsection{Independent Variable: Injection Strategy}
\label{sec:iv}

\begin{description}
    \item[\none:] The \agentsmd{} file is removed from the workspace. No repository context is provided in the system prompt or accessible to the agent. The agent works from the codebase alone.
    \item[\alwayson:] The full \agentsmd{} content is injected into the system prompt every turn, preceded by a framing sentence (``The following is the repository's AGENTS.md guide...'').  The file is removed from the workspace to prevent double-reading.
    \item[\selective:] Topic-organized wiki files (e.g., \texttt{wiki/architecture.md}, \texttt{wiki/error-handling.md}) are placed in the workspace, and a system-prompt hint tells the agent to consult \texttt{wiki/*.md}; the agent retrieves relevant files on demand via its Read tool. \textbf{Caveat on the \selective{} corpus.} The wiki is content-matched to the \agentsmd{} for only one repository (opshin: a single \texttt{overview.md} equal to the 256-word file). For pdm and firebase the wiki is a broader, auto-generated repository wiki---roughly $10\times$ and $18\times$ the words of the \agentsmd{}, respectively---so for those repositories \selective{} varies \emph{both} the delivery channel \emph{and} the context corpus. We treat this as a known confound: it means \selective{} is best read as ``on-demand retrieval from a repository wiki'' rather than a pure re-packaging of the \agentsmd{}. Notably, the larger corpus only strengthens the correctness null---it gave the agent strictly more material and still did not raise pass-rates (\S\ref{sec:results-correctness}).
\end{description}

In all conditions, the \agentsmd{} file itself is removed from the workspace, so each strategy delivers only what it explicitly provides.
The within-strategy contrasts are clean; the corpus caveat above qualifies the \selective{} arm and is revisited in the Limitations.

\subsection{Agents}

We run two frontier coding agents from different providers:
\begin{itemize}
    \item \textbf{Claude Code} (\texttt{claude-sonnet-4-6}, Anthropic). Multi-turn, tool-use agent invoked via \texttt{claude -{}-print -{}-bare -{}-output-format stream-json}. Context injected via \texttt{-{}-append-system-prompt}. Cost-capped at \$6/run.
    \item \textbf{Codex CLI} (\texttt{gpt-5.5}, OpenAI, ChatGPT-plan auth). Single-session agent invoked via \texttt{codex exec -{}-json}. Context injected by prepending to the user prompt (Codex lacks a system-prompt flag). Turn-capped at 120 via watchdog.
\end{itemize}

Both agents use the same workspace, task prompt, and evaluation pipeline.
The injection-channel asymmetry (system prompt vs.\ user-turn prepend) is a noted confound (see Limitations).

\subsection{Efficiency Metrics}

We distinguish \emph{portable} metrics (comparable across agents) from \emph{agent-specific} metrics:
\begin{itemize}
    \item \textbf{Portable:} tool calls, wall-clock duration, output tokens.
    \item \textbf{Agent-specific:} cache-read tokens, cache-creation tokens (Claude only; Codex folds cached input into \texttt{input\_tokens} with no separate accounting).
\end{itemize}

Turns are excluded as a cross-agent metric: Codex emits exactly 1 \texttt{turn.completed} event per session regardless of work performed (a classifier artifact we discovered during screening; \S\ref{sec:classifier}).

\subsection{Statistical Analysis}

The unit of analysis is the \emph{task} (not the repeat).
For each (task, strategy) pair, the 3 repeats are averaged to yield one value per task per strategy.
We then apply:
\begin{itemize}
    \item \textbf{Omnibus permutation test:} strategy labels permuted within-task (10,000--20,000 iterations); test statistic = variance of marginal strategy pass-rates.
    \item \textbf{Paired Wilcoxon signed-rank tests} with Holm-Bonferroni correction for efficiency metrics (family of 12 tests across 3 pairs $\times$ 4 metrics).
    \item \textbf{TOST equivalence:} task-clustered bootstrap (10k) on paired strategy differences; equivalence declared if 95\% CI falls within $\pm\delta$.
    \item \textbf{Monte Carlo power:} simulate binary outcomes under varying true $\Delta$ and $n_{\text{tasks}}$ to determine minimum detectable effect (MDE) and sample size for 80\% power.
\end{itemize}

\section{Results}

\subsection{Correctness: No Strategy Effect (Both Agents)}
\label{sec:results-correctness}

Table~\ref{tab:correctness} presents pass-rates by strategy and agent.
Neither agent shows a statistically significant strategy effect.

\begin{table}[t]
\centering
\small
\caption{Pass-rate by strategy and agent. Omnibus permutation $p$-values test whether strategy explains correctness beyond task difficulty.}
\label{tab:correctness}
\begin{tabular}{lcc}
\toprule
\textbf{Strategy} & \textbf{Claude (15 tasks)} & \textbf{Codex (17 tasks)} \\
\midrule
\none        & 53.3\% (24/45) & 58.8\% (30/51) \\
\alwayson    & 55.6\% (25/45) & 56.9\% (29/51) \\
\selective   & 55.6\% (25/45) & 52.9\% (27/51) \\
\midrule
Omnibus $p$  & 1.000          & 0.66 \\
\bottomrule
\end{tabular}
\end{table}

For Claude, all pairwise strategy differences are ${\leq}2.3$pp.
For Codex, the largest difference is 5.9pp (\none{} vs.\ \selective{}), driven by per-task noise at $n{=}3$ repeats.
The omnibus permutation test shows no detectable strategy effect ($p{=}1.00$ Claude; $p{=}0.66$ Codex).
We note this test is intrinsically low-power here: the floor/ceiling structure (\S\ref{sec:borderline}) leaves little marginal-pass-rate variance to permute, so the $p{=}1.00$ is close to mechanical and the substantive null is carried by the dynamic-range (borderline) subset analyzed below, not by the omnibus alone.

\begin{figure*}[t]
\centering
\includegraphics[width=0.92\textwidth]{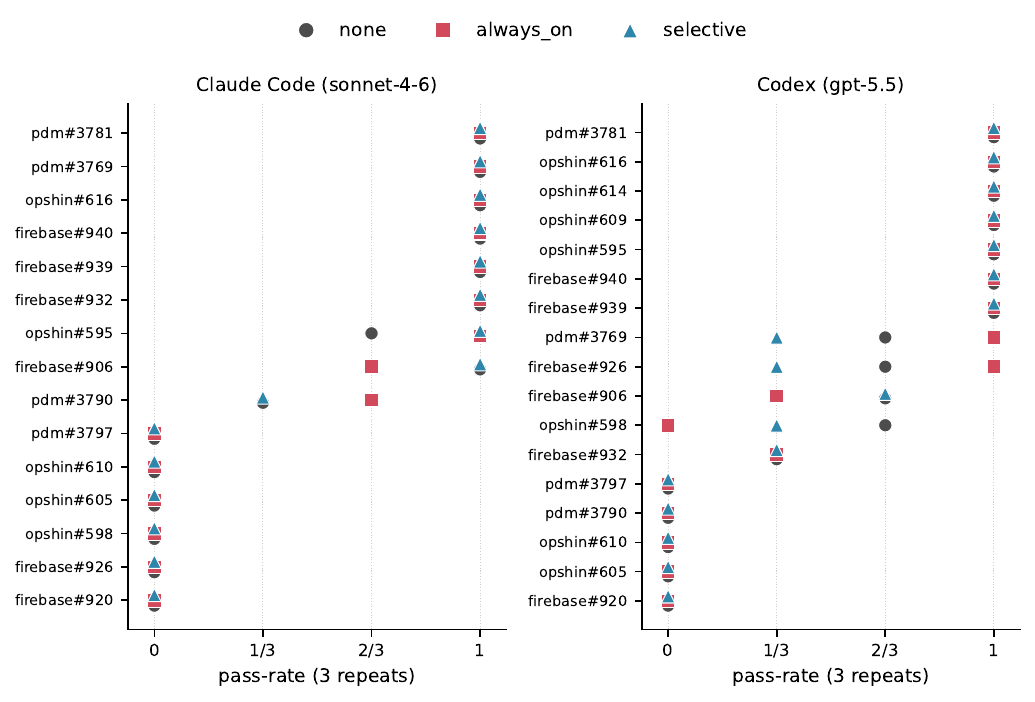}
\caption{Per-task pass-rate (3 repeats) by injection strategy, one panel per agent, tasks ordered by mean pass-rate. For the large majority of tasks the \none{}/\alwayson{}/\selective{} markers coincide---strategy does not move correctness---and tasks sit at the floor (0) or ceiling (1). The markers separate only on the handful of borderline tasks, which differ by agent (e.g.\ pdm\#3790 for Claude; pdm\#3769, firebase\#926 for Codex), illustrating the agent-specific difficulty of \S\ref{sec:borderline}.}
\label{fig:pertask}
\end{figure*}

Figure~\ref{fig:pertask} makes the null visible at the task level: for almost every task the three strategy markers coincide at a pass-rate of 0 or 1 (the floor/ceiling structure of \S\ref{sec:borderline}), and the few tasks where the markers separate are precisely the agent-specific borderline tasks---different tasks for each agent.

\paragraph{The null holds with dynamic range.}
A potential objection is that the null reflects floor/ceiling effects (tasks too easy or too hard for context to matter).
We address this with borderline-task expansion: on 4 Codex-borderline tasks (17--67\% baseline pass rate), \none{} achieves 58\% vs.\ \alwayson{} 42\% and \selective{} 42\%---context injection does \emph{not} help even where the design has power to detect an effect.

\paragraph{Equivalence bounds.}
Descriptive equivalence testing (TOST on the task-clustered bootstrap) bounds every pairwise strategy difference to $<$10pp for Claude and $<$15pp for Codex---compatible with the $>$30pp MDE (\S\ref{sec:power}) because the observed point estimates are ${\approx}0$, not a powered equivalence claim given $n{=}15$--$17$ clusters.

\subsection{Efficiency: Two Narrow Process Signals, Otherwise Null}

\paragraph{Claude.}
\selective{} uses significantly less cache-creation tokens than \none{} (unanimous 11/11 tasks lower; $p{=}0.001$, $p_{\text{Holm}}{=}0.012$).
Directional but non-significant trends appear for cache-read (9/11 lower, $p{=}0.067$) and \alwayson{} vs.\ \selective{} duration (8/11 lower, $p{=}0.054$).
No significant effect on turns or tool calls after correction.
We read this cache result mechanically rather than as a strategy benefit: \selective{} keeps only a short retrieval hint in the system prompt and reads a few files on demand, whereas \alwayson{} re-presents the whole \agentsmd{} each turn---so the footprint difference follows from the delivery mechanics (and, for pdm/firebase, from a different wiki corpus; \S\ref{sec:iv}), not from context making the agent more capable.

\paragraph{Codex.}
All efficiency metrics are flat across strategies: tool calls 32/32/32, output tokens $\pm$3.8\%, duration $\pm$3.8\% (all $|d_z|{<}0.2$, negligible effect size).

\paragraph{A within-agent process effect (Claude\,$\times$\,opshin).}
A second, narrow efficiency signal survives. On \textbf{opshin}---the one repository whose \agentsmd{} carries explicit \emph{runtime/compile-time} warnings (``the full test suite takes $>$20 minutes'')---Claude's within-task wall-clock time is ${\sim}$24\% lower under context (\none{} $2689$s vs.\ \alwayson{} $2066$s vs.\ \selective{} $2032$s; faster-with-context on 4/5 tasks, sign-flip $p{=}0.125$, underpowered at $n{=}5$).
What lifts this above noise is a \emph{dose-dependent} mechanism: counting the agent's \texttt{pytest} invocations, the number of \emph{blind full-suite} runs per cell falls monotonically \none{} $3.67 \to$ \alwayson{} $2.44 \to$ \selective{} $1.67$ (Table~\ref{tab:opshin}), directional under a within-task paired test (fewer on 3/4 tasks, $p{=}0.25$, $n{=}4$; per-task counts in Appendix~\ref{app:pertask}).
Stripped of the \agentsmd{} warning, the agent repeatedly runs the slow full suite; given the warning, it runs targeted tests instead.
The dose ordering (\selective{}\,$<$\,\alwayson{}\,$<$\,\none{}) matches both the full-suite count and the duration.
The scope is deliberately narrow: the effect appears only for Claude (Codex duration is flat) and only on opshin (firebase shows the opposite ${+}82$s direction---a fast repository with little to warn about).
Like the cache signal, it is a \emph{process} effect (how the agent works), not an \emph{outcome} effect (correctness, still null).
We report it as \emph{exploratory}: it emerged from a post-hoc mechanism analysis rather than the pre-registered efficiency-metric family, and we therefore do not fold it into the Holm-corrected confirmatory tests below.

\begin{table}[t]
\centering
\small
\caption{Claude on opshin: context reduces blind full-suite test runs and wall-clock time. Full-suite count = \texttt{pytest} invocations targeting the whole suite per cell.}
\label{tab:opshin}
\begin{tabular}{lcc}
\toprule
\textbf{Strategy} & \textbf{Full-suite runs/cell} & \textbf{Duration (s)} \\
\midrule
\none      & 3.67 & 2689 \\
\alwayson  & 2.44 & 2066 \\
\selective & 1.67 & 2032 \\
\bottomrule
\end{tabular}
\end{table}

\paragraph{Cross-agent comparison (portable metrics, qualitative).}
Codex is consistently leaner (fewer tool calls, less wall-clock time, fewer output tokens) at a slightly lower pass-rate.
We report this only qualitatively: the agents differ in injection channel (\S\ref{sec:iv}), tokenizer, and token accounting (Codex folds cached input into \texttt{input\_tokens}), so any between-agent gap is confounded. Our actual contrast---the \emph{within}-agent strategy comparison---is unaffected.

\subsection{Agent-Specific Borderline Tasks}
\label{sec:borderline}

A striking cross-agent finding: the \emph{same task} occupies different difficulty bands depending on the agent.

\begin{table}[t]
\centering
\small
\setlength{\tabcolsep}{5pt}
\caption{Agent-specific difficulty. Pass counts none/always/selective; \textbf{bold} marks the borderline outcome. pdm\#3790 is borderline for Claude only; pdm\#3769 for Codex only.}
\label{tab:borderline}
\begin{tabular}{@{}lcc@{}}
\toprule
\textbf{Task} & \textbf{Claude} & \textbf{Codex} \\
\midrule
pdm\#3790 & 1/3 · \textbf{2/3} · 1/3 & 0/3 · 0/3 · 0/3 \\
pdm\#3769 & 3/3 · 3/3 · 3/3 & 2/3 · \textbf{3/3} · 1/3 \\
\bottomrule
\end{tabular}
\end{table}

This is not merely anecdotal. Across the 15 tasks both agents attempted, per-task pass rates are positively but imperfectly correlated (Spearman $\rho{=}0.75$, $p{=}0.001$; Pearson $r{=}0.77$): difficulty largely transfers, yet the \emph{informative band} does not.
Of the 15 shared tasks, 6 are borderline ($0{<}\text{pass}{<}1$) for exactly one agent, and 6 differ in floor/ceiling status across agents---so for roughly 40\% of tasks, the agent that would reveal a context effect is not the same agent.

This offers a candidate explanation for prior contradictions---a hypothesis we cannot test directly without their exact task sets: if Paper 1 (Codex-family) and Paper 2 (Claude-family) drew tasks that landed in different agents' borderline bands, they would observe different context effects---not because context inherently helps or doesn't, but because the informative range is agent-specific.
A practical consequence is that ablation studies must screen tasks \emph{per agent}; a borderline set calibrated on one agent gives another agent mostly floor/ceiling tasks, where no manipulation can show an effect.

\subsection{Classifier Portability Lesson}
\label{sec:classifier}

During borderline screening, our effort classifier (split on turns${\geq}$30) silently misclassified all Codex tasks as ``trivial'' because Codex emits exactly 1 \texttt{turn.completed} event per session.
Eight genuinely high-effort tasks (18--51 tool calls, 410--1220s) were incorrectly dropped.
Reclassifying on \emph{tool calls}---the pre-registered portable metric---recovered them.
This demonstrates that turn-based metrics are non-portable across agent architectures.

\subsection{Power Analysis and Detection Limits}
\label{sec:power}

Our Monte Carlo simulation (\S3.6) reveals:
\begin{itemize}
    \item MDE at $n{=}17$, reps$=$3: even a large $\Delta{=}30$pp effect is caught only 57\% of the time; smaller effects fare worse.
    \item Detecting $\Delta{=}10$pp at 80\% power requires ${\sim}$120--200 tasks.
    \item Adding repeats barely helps ($n{=}17$, reps $3{\to}10$: power increases 13\%${\to}$58\% for $\Delta{=}15$pp).
\end{itemize}

Task-level variance dominates: scaling requires more \emph{tasks}, not more repeats.
This is a methodological finding relevant to future ablation study designs.

\section{Discussion}

\subsection{Why the Null Holds: Skill, Not Knowledge}

The failure-mode triage (\S\ref{sec:triage}) reveals \emph{why} context files do not help.
We inspected the near-miss failures (1--4 failing gold tests)---the tasks most likely to flip with one additional piece of information:

\begin{itemize}
    \item \textbf{Task 510 (opshin):} requires a subtle union-expansion optimization. The agent built the entire optimization pass but introduced a correctness bug. Gap: engineering precision, not a missing fact.
    \item \textbf{Task 907 (firebase):} requires proactive auth-token refresh. The agent implemented reactive retry instead. Gap: architectural pattern choice, not secret knowledge.
    \item \textbf{Task 554 (opshin):} requires rejecting V2 validator arguments. The agent knew the V3 rule (from the code) but miswired the check. Gap: exact behavioral specification.
    \item \textbf{Task 593 (opshin):} requires type-narrowing through \texttt{isinstance} + \texttt{assert}. Gap: deep type-system reasoning.
\end{itemize}

None of these failures isolate a knowable-fact gap that an \agentsmd{} could fill.
They fail on \emph{implementation skill}---feature design, pattern selection, exact wiring---not on missing repository-private knowledge.
\label{sec:triage}

\subsection{Manipulation-Validity Probe}
\label{sec:probe}

The triage motivates an obvious objection: perhaps the null reflects an \emph{inert manipulation}---we injected text that did nothing measurable.
We address this directly with a pre-registered probe.
First, we rated our three context files against a structured quality rubric we applied across a 40-repository \agentsmd{} corpus (scoring coverage of build/test/style conventions, architectural guidance, and specificity); on this rubric \emph{firebase} rates ``Excellent,'' \emph{pdm} and \emph{opshin} ``Good''---they are not low-quality files. This is our own rubric assessment, not a third-party audit.
Second, we re-ran the two convention-closest near-misses (opshin~554, firebase~907) under all three strategies, 3 repeats, on \emph{both} agents (36 cells), to test whether the real, unmodified \agentsmd{} can flip a near-miss to a pass (Table~\ref{tab:probe}).

\begin{table}[t]
\centering
\small
\caption{Manipulation-validity probe: pass-rate (none/always/selective) on the two convention-closest near-miss tasks. The real \agentsmd{} never converts a failure to a pass on either agent; in the single task with cross-agent range (Claude, 907), more context does not improve correctness.}
\label{tab:probe}
\begin{tabular}{lcc}
\toprule
\textbf{Task} & \textbf{Codex} & \textbf{Claude} \\
\midrule
907 (firebase) & 0/3 · 0/3 · 0/3 & \textbf{2/3} · 1/3 · 0/3 \\
554 (opshin)   & 0/3 · 0/3 · 0/3 & 0/3 · 0/3 · 0/3 \\
\bottomrule
\end{tabular}
\end{table}

The result confirms the pre-registration and sharpens it.
(1)~\emph{No helpful flip anywhere}: the real \agentsmd{} never rescues a near-miss (Codex fails 18/18 regardless of strategy; task~907's \none{} near-miss of 109~pass / 1~fail never crosses).
(2)~\emph{In the one task with cross-agent range, more context did not help}: task~907 is Codex-hard but Claude-borderline, and Claude passes it 2/3 under \none{}, 1/3 under \alwayson{}, 0/3 under \selective{}. We do not claim a general downward effect---this is a single task at $n{=}3$ repeats---but the trend is clearly non-positive, and no condition on either agent produced a flip toward passing.
(3)~Task~554 is Claude-floored, another instance of agent-specific difficulty (\S\ref{sec:borderline}).
The probe is a stronger answer to ``did your manipulation do anything?'' than a bare null: across both agents the real \agentsmd{} never converts a near-miss to a pass.
The manipulation is not inert (it can perturb behavior, and our files rate Good/Excellent on our quality rubric); it simply does not supply the implementation skill that gates these tasks.

\subsection{Reconciling Prior Work}

Our results suggest a reconciliation of the Paper 1 / Paper 2 contradiction (offered as a hypothesis, since we do not have access to their exact task sets):
\begin{enumerate}
    \item Context files \emph{do not} improve correctness on either agent family (confirming Paper 2).
    \item The efficiency gain Paper 1 reports likely reflects the \emph{mechanical} cost of context injection (larger prompts $\to$ more tokens), not a task-performance improvement. Notably, Paper 2 reports the opposite sign on its own cost metric---context files \emph{increase} inference cost by over 20\%---consistent with injection mechanics, not agent capability, driving these numbers in either direction depending on setup. Our \selective{} strategy achieves a real cache-efficiency reduction (p$_\text{Holm}$=0.012), but this is likewise an artifact of injection mechanics, not agent capability.
    \item Prior disagreements may reflect agent-specific borderline effects (\S\ref{sec:borderline}): a task in one agent's informative band appears in another's floor/ceiling.
\end{enumerate}

\subsection{Practitioner Implications}

The \agentsmd{} files in our sample are real-world context files representative of practitioner effort---style guides, architectural notes, error-handling conventions.
Scoped to our setting---3 Python repositories, naturalistic style-guide-type context, and the injection \emph{channel} rather than content---we detect no correctness benefit: \textbf{in this sample, generic context files do not measurably improve coding-agent correctness} (bounded to ${<}10$--$15$pp; see Limitations).
We cannot rule out benefits for other languages, larger repositories, or purpose-built task-specific context; the safe reading is that effort spent on \emph{generic} context documents may pay off less than effort on task decomposition, tooling, or example-driven prompting.
One nuance is genuinely actionable, however: even when context does not change pass/fail, it can change \emph{how} the agent works---our \selective{} strategy measurably reduces wasted full-suite test runs (Table~\ref{tab:opshin})---so context may still earn its place on cost and latency grounds.

\section{Conclusion}

We presented a controlled ablation of context-injection strategy for coding agents, spanning 291 agent runs (288 evaluated) across two frontier agents and 17 real-world tasks, plus a 36-cell manipulation-validity probe.
Context-injection strategy does not measurably move correctness (observed effects bounded to $<$10--15pp via descriptive TOST equivalence; not a powered equivalence claim, \S\ref{sec:power}).
The two surviving efficiency effects are both narrow and \emph{process-level}: a cache-footprint reduction for Claude's \selective{} strategy, and a dose-dependent reduction in blind full-suite test runs on the one repository whose context file warns about test cost---neither changes the correctness outcome.

Our failure-mode analysis and probe reveal the mechanism: real coding tasks fail on implementation skill, not missing repository knowledge that a context file could supply; across both agents the real context file never converts a near-miss to a pass (and in the single Claude task with cross-agent dynamic range, at $n{=}3$ repeats, the trend was non-positive---a single-task observation, not a general downward effect).
The ``borderline is agent-specific'' finding explains prior contradictions and implies that future ablation studies must screen tasks per-agent to achieve informative dynamic range.

We release our full experimental harness, 291-run dataset (288 evaluated cells), and power analysis code to support future work on scaling this design to the ${\sim}$120-task threshold needed for powered equivalence at 10pp.

\section*{Limitations}
\label{sec:limitations}

\begin{enumerate}
    \item \textbf{Sample size.} 15/17 tasks with 3 repeats. MDE $>$30pp; a 10pp effect is undetectable. Our TOST bounds the effect to $<$10--15pp but cannot achieve narrower equivalence without $\sim$120 tasks.
    \item \textbf{Repository diversity.} Three Python repositories. Results may not generalize to other languages, larger codebases, or repositories with exceptionally detailed context files.
    \item \textbf{Injection-channel asymmetry.} Claude receives context via system prompt; Codex via user-turn prepend (no system-prompt flag). This is a confound between the agent arms, though the within-agent strategy comparison remains clean.
    \item \textbf{Ecological validity.} Our \alwayson{} condition injects context via system prompt every turn, which is stronger than the natural workflow (agent reads the file once from the workspace). We argue that if guaranteed presence does not help, natural discovery cannot either---but this is an inference, not a direct measurement. A fourth ``natural'' condition where the file simply exists in the workspace would strengthen ecological validity.
    \item \textbf{Selective is our construction, and its corpus is not content-matched.} The wiki-split design and retrieval hint are specific choices; alternative selective strategies (e.g., semantic retrieval) might yield different results. More importantly, the \selective{} wiki equals the \agentsmd{} only for opshin; for pdm and firebase it is a broader auto-generated repository wiki ($\sim$10$\times$/18$\times$ larger), so \selective{} confounds delivery channel with context corpus for two of three repositories (\S\ref{sec:iv}). This does not threaten the correctness null (the larger corpus, if anything, gave \selective{} an advantage it did not convert into passes), but it does mean we cannot cleanly attribute the \selective{} cache-footprint reduction to channel alone, and a content-matched split is required before interpreting \selective{} as a pure channel manipulation. We leave an equal-content \selective{} arm across all repositories to future work.
    \item \textbf{Inert-manipulation concern.} Our context files are naturalistic (not purpose-built for specific tasks). The manipulation-validity probe (\S\ref{sec:probe}) shows the real \agentsmd{} \emph{can} perturb behavior but never converts a near-miss to a pass, and our own rubric assessment rates the files Good/Excellent---so the null is not an artifact of low-quality or inert context. Whether \emph{purpose-built, task-specific} context (a fact the agent provably cannot infer) would help remains an open question for future work.
    \item \textbf{Model-version snapshot.} All results are specific to \texttt{claude-sonnet-4-6} and \texttt{gpt-5.5} as of this study; agent behavior, and hence the null, may shift as these models are updated.
    \item \textbf{Mixed provenance.} Claude repeat-0 ran on a local machine; repeats 1--2 on the pod. A sensitivity analysis that drops repeat-0 entirely and re-estimates the contrasts from the pod-only repeats leaves the correctness null intact: Claude's marginal pass-rates remain 53--55\% under all three strategies, and every paired strategy contrast stays within ${\pm}3.3$pp (largest shift: \selective{}\,$-$\,\none{} moves from $+2.2$ to $0.0$pp). Behavior is also stable on matched cross-check cells (e.g.\ 22 vs.\ 23 turns). The provenance split therefore does not drive the result.
\end{enumerate}

\section*{Ethics Statement}
This study involves no human subjects or personal data. The experimental subjects are
AI coding agents operating on public open-source repositories. All agent runs execute in
an egress-locked sandbox that prevents any modification of external state (no pushes,
commits, or pull requests reach GitHub); see Appendix~\ref{app:harness}.
AI coding agents (Claude Code, Codex) are the \emph{object of study}, not authoring tools.
Separately, the authors used an AI assistant for drafting support and code scaffolding;
all experimental results, statistics, and claims were verified by the authors against the
released data.

\bibliography{references}

\appendix
\section{Experimental Harness Details}
\label{app:harness}

\paragraph{Safety layers.}
Our harness implements defense-in-depth to prevent the agent from modifying external state:
(1) DNS blackhole for GitHub (egress lock);
(2) \texttt{git remote remove} on all workspace remotes;
(3) \texttt{git push/commit/remote} denied via PATH shims and Claude's \texttt{-{}-disallowedTools};
(4) \texttt{GH\_TOKEN}/\texttt{GITHUB\_TOKEN} stripped from environment;
(5) future git history pruned (agent cannot \texttt{git show} the gold commit).

\paragraph{Evaluation pipeline.}
After agent completion: (1) capture diff against \texttt{base\_sha};
(2) restore gold test files to base state;
(3) apply gold test patch;
(4) run \texttt{pytest} with per-repo configuration;
(5) pass iff all gold tests pass with zero failures/errors.

\section{Per-Task Results}
\label{app:pertask}

Figure~\ref{fig:pertask} plots per-task pass-rates for both agents. The complete per-task pass/fail counts are available in the supplementary material (\texttt{results\_summary.csv}) alongside the \texttt{experiment\_full.db} SQLite database.

\paragraph{Opshin full-suite test runs (\S\ref{sec:results-correctness}).}
Per-task blind full-suite \texttt{pytest} counts (\none{}\,/\,mean-context) behind Table~\ref{tab:opshin}: task~595 $2.50/0.75$, task~605 $1.20/1.25$, task~610 $2.50/0.67$, task~616 $9.33/6.25$. Three of four tasks run fewer full-suite invocations under context; task~605 runs counter to the trend.

\section{Reproduction and Data Availability}
\label{app:reproduce}

We release the full experimental harness, the context-injection strategy
implementations (\none{}/\alwayson{}/\selective{}), the defense-in-depth safety
layer, the task specifications, the statistical analysis code, the 291-run
ablation dataset (288 evaluated cells) and 36-cell probe (\texttt{experiment\_full.db},
\texttt{probe\_codex.db}, \texttt{probe\_claude.db}), and an aggregated per-cell
results table (\texttt{results\_summary.csv}).
Raw agent transcripts are available from the authors on request.
The repository is public: \url{https://github.com/codeprakhar25/context-files-coding-agents}.

{\footnotesize
\begin{verbatim}
# Requires: experiment_full.db
python3 power_analysis.py \
        experiment_full.db codex
python3 power_analysis.py \
        experiment_full.db claude_code
python3 efficiency_stats_correct.py
\end{verbatim}
}

\end{document}